# Theoretical, Measured and Subjective Responsibility in Aided Decision Making

NIR DOUER, Tel Aviv University

JOACHIM MEYER, Tel Aviv University

When humans interact with intelligent systems, their causal responsibility for outcomes becomes equivocal. We analyze the descriptive abilities of a newly developed responsibility quantification model (ResQu) to predict actual human responsibility and perceptions of responsibility in the interaction with intelligent systems. In two laboratory experiments, participants performed a classification task. They were aided by classification systems with different capabilities. We compared the predicted theoretical responsibility values to the actual measured responsibility participants took on and to their subjective rankings of responsibility. The model predictions were strongly correlated with both measured and subjective responsibility. A bias existed only when participants with poor classification capabilities relied less-than-optimally on a system that had superior classification capabilities and assumed higher-than-optimal responsibility. The study implies that when humans interact with advanced intelligent systems, with capabilities that greatly exceed their own, their comparative causal responsibility will be small, even if formally the human is assigned major roles. Simply putting a human into the loop does not assure that the human will meaningfully contribute to the outcomes. The results demonstrate the descriptive value of the ResQu model to predict behavior and perceptions of responsibility by considering the characteristics of the human, the intelligent system, the environment and some systematic behavioral biases. The ResQu model is a new quantitative method that can be used in system design and can guide policy and legal decisions regarding human responsibility in events involving intelligent systems.

• **Applied computing → Operations research→** *Decision analysis* • **Mathematics of computing → Information theory**

## 1 INTRODUCTION

Intelligent systems, which can perceive, analyze and respond to the world around them, have become major parts of our everyday lives. They are prominent in factory automation (e.g., automated production facilities and advanced control rooms), transportation (e.g., aviation and autonomous vehicles), medical care (e.g., advanced data-based decision support systems), military applications (e.g., intelligence systems and autonomous weapon systems), and many other domains, such as entertainment, e-commerce, assistive robotics and more. In these systems, computers and humans share the collection and evaluation of information, decision-making and action implementation.

In the interaction with such intelligent systems, the human comparative responsibility becomes equivocal. For instance, what is the human responsibility when all information about an event arrives through a system that collects and analyzes data from multiple sensors, without the human having any independent information? If a human performs an action the system indicated as necessary, is the human responsible for the action, if it caused harm? The determination of the human causal responsibility is critical in the design and investigation of intelligent systems that can lead to injury and even death, such as autonomous vehicles, automated use of hazardous materials in industry, or autonomous weapon systems.

We developed a theoretical Responsibility Quantification model (the ResQu model) of human responsibility in intelligent systems [Douer and Meyer 2020]. The ResQu model provides for the computation of a responsibility measure to quantify human causal responsibility in interactions with intelligent systems. This theoretical measure predicts the average share of unique human contribution to the overall outcomes, based on different characteristics of the operational environment, the system and the human, and the allocation of functions between them.

The ResQu model is a normative model. It assumes perfect rationality on the part of the human, perfect knowledge about probabilities and properties of the system and optimal human utilization of the system. However, in reality, people may act non-optimally when they interact with intelligent systems [Alvarado-Valencia & Barrero 2014, Arnott 2006, Baker et al. 2004, Goddard et al. 2011, Mosier et al. 1998, Parasuraman & Riley 1997]. This raises the question whether this normative model can also serve as a descriptive model for predicting actual human behavior with intelligent systems.

We address this question here and report two controlled experiments in which participants interacted with a simple decision support system in the controlled settings of a laboratory. Based on the observed behavior, we computed the measured responsibility, defined as the observed average share of unique human contribution to the outcomes. We compared these scores to the corresponding theoretical ResQu predictions for the different experimental conditions.

Responsibility is also a psychological phenomenon. People may perceive their contribution to a process differently from their actual contribution. Thus, it is also important to analyze the relation between subjective perceptions and both theoretical and measured responsibilities. To do so, we asked participants to evaluate their contributions. We compared the subjective evaluations to the corresponding theoretical and measured responsibility values. Fig. 1 summarizes the descriptions and properties of the three responsibility measures we use in this paper.

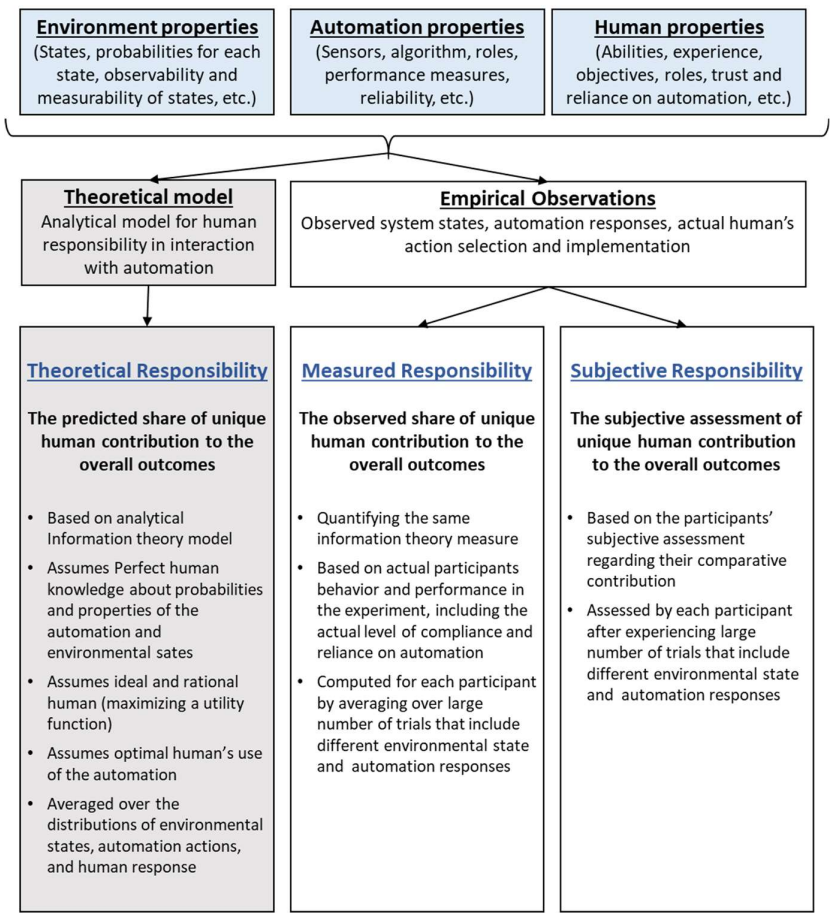

Fig. 1. Descriptions and properties of theoretical, measured, and subjective responsibility

## 2 BACKGROUND AND RELATED WORK

### 2.1 Human Responsibility in the Interaction with Intelligent Systems

Philosophical and legal research includes extensive studies on the concept of responsibility, investigating different aspects such as role responsibility, causal responsibility, liability (or legal responsibility) and moral responsibility [Hart & Honor 1985, Hart 2008, Vincent 2011]. In the domain of human interaction with intelligent systems, *role responsibility* refers to assigning specific roles and duties to the human operator for which the human is accountable to others. However, this role assignment does not specify the causal relations between the human's actions and the overall consequences and outcomes of the human interaction with the system. This relation is better defined by *causal responsibility*, which describes the actual human contribution to the combined human-machine outcomes.

The ability to control a system and the resulting consequences is a necessary condition for assigning causal responsibility [Noorman 2014]. So far, causal responsibility was usually associated with humans. When an event involved technology, the responsibility was usually with the user, unless some unexpected circumstances existed. Manufacturers of systems could also be held responsible if, for instance, their product failed to fulfill legal requirements or standards. With the introduction of intelligent systems, a shift occurred towards shared control, in which the human and computerized systems jointly make decisions or control actions [Abbink et al. 2018]. These are combined to generate a final control action or decision. There may also be supervisory control, in which the human sets high-level goals, monitors the system and only intervenes if necessary. Moreover, in advanced systems, which incorporate artificial intelligence, machine-learning and neural networks, developers and users may be unable to fully control or predict all possible behaviors and outcomes, since their internal structure can be opaque (a "black box") and sometimes can yield odd and counterintuitive results [Castelvecchi 2016, Scharre 2016]. Consequently, humans may no longer be able to control intelligent systems sufficiently to be rightly considered fully responsible for their outcomes. The intelligent system (or its developers) may share some of the responsibility [Johnson & Powers 2005, Coeckelbergh 2012], similar to the legal concept of comparative responsibility, a doctrine of tort law that divides fault among different parties [Cooter & Ulen 1986, Pinto 1978, Sobelsohn 1984]. This difficulty to determine human causal responsibility when using advanced intelligent systems has created a "responsibility gap" in the ability to divide causal responsibility between humans and intelligent systems [Docherty et. al. 2012, Johnson 2014, Matthias 2004].

The responsibility gap is clearly exhibited in human interaction with advanced decision support systems (DSS). Decision support systems are information systems that support and improve the human decision making and action selection processes. They have become vital tools in many domains [Arnott & Pervan 2005, Turban et al. 2005]. With the advent of AI, these systems progressed into intelligent decision support systems (IDSS), which incorporate AI technologies, such as machine learning, big-data analytics, fuzzy logic, neural networks, genetic algorithms etc. [He & li 2017, Turban et al. 2005]. IDSS are increasingly used in fields, such as medical diagnostics [AlSalman & Almutairi 2019, Contreras & Vehi 2018, Dhombres et al. 2019, Elalfi et al. 2016, Hung et al. 2019, Tan et al. 2016], industrial applications [Agrawal et al. 2016, Cavalieri 2004, Irannezhad 2020, Rodríguez 2019, Rikalovic 2017, Sellak et al. 2017] and web-based applications, such as intelligent agents and recommendation systems, which are widely used for e-commerce, targeted advertising, consumer segmentation, information search and indexing and more [Turban et al. 2005]. Using AI technologies, IDSS take over large parts of processes that relied so far on human decision-making, including the acquisition of information, its analysis, and the selection of actions. There may be cases at which a user cannot knowledgably supervise the opaque AI processes or must make decisions, based exclusively on the recommendations from the IDSS, without being able to evaluate them independently. Hence, it is difficult to determine human causal responsibility in interactions with IDSS.

The above difficulty is even greater with mixed initiative systems, which are a special kind of intelligent decision support systems. In mixed initiative systems, not only the human operator initiates the human-machine interactions, but the system can also initiate an action, involving the human user. Mixed initiative systems proactively contribute to the interaction and solution process through an iterative evolution of interactions between humans and machines that are often seen as a conversation between the two parties [Liapis & Shaker 2016].There are various initiatives the system can share, such as task initiative (suggesting what problem needs to be solved now), speaker initiative (determining when each actor will speak or may interrupt the other) and outcome initiative (deciding how the selected problem should be solved) [Novick & Sutton 1997]. To do so, these systems combine AI for inferring the user's intentions and focus, employ dialog to resolve remaining uncertainties they have regarding the user, use context-dependent costs and benefits, consider deferring actions to a time when they will be less distracting to the user, and learn continuously to become a better teammates for their user [Horvitz 1999]. With mixed initiative systems, humans and systems not only act together to create an outcome (co-create), but the system's initiatives may directly foster new human creativity and actions [Yannakakis et al. 2014]. Hence, with mixed initiative systems, it is particularly difficult to determine the comparative human causal responsibility for outcomes.

The equivocal nature of human responsibility in interactions with intelligent systems is not only an abstract, philosophical question. The rapid development of intelligent systems have raised concerns that humans will become less and less involved in their use, and thus, they will be considered or feel less responsible for adverse outcomes [Crootof 2015, Cummings 2006]. This prompted the demand to involve humans in automated processes in a manner that will facilitate meaningful human control [de Sio & Van den Hoven 2018]. This is the case, for example, with advanced weapon systems, in which the issue of "meaningful human control" has become a key topic in discussions [Crootof 2016, Heyns 2016, Horowitz & Scharre 2015, Neslage 2015, UNIDIR 2014]. The demand to facilitate meaningful human control, has also surfaced in other intelligent systems, such as computers, surgical robotics, autonomous cars, and more [de Sio, F & Van den Hoven, 2018, Ficuciello et al. 2019, Mecacci & de Sio 2019).

However, simply putting a human into the loop does not assure that the human involvement will be meaningful. When system's capabilities greatly exceed those of the human, the human may not be able to knowledgably supervise the system or provide significant contribution. Currently, there are different, and sometimes contradicting interpretations and policies regarding meaningful human involvement, and system designers lack models and metrics to address this issue systematically [Canellas & Haga 2015].

The theoretical ResQu model we developed [Douer and Meyer 2020], aimed to provide measures of the division of causal responsibility between humans and intelligent systems. Using information theory, it defines a method to compute the expected share of the unique human contribution to the overall outcomes, based on the characteristics of the operational environment, the system and the human, and the function allocation between them. In addition, it can be used to quantify the level of meaningful human control, based on the premise that meaningful human control requires the human to have some causal responsibility for outcomes.

## 2.2 The ResQu Model formulation for Binary Classification Systems

In the current study we examined the ability of the ResQu model to describe human behavior in the interaction with decision support systems that perform binary classifications. Binary classification systems, such as binary alerts, warn the user about abnormal conditions or about measured values that exceed some threshold. These are the simplest form of intelligent systems. They are the most widely used decision aid, installed in flight decks, industrial control rooms, vehicles, medical equipment, computer-aided diagnostic systems, smart homes, and many other computerized and AI systems [Bregman 2010, Cicirelli et al. 2016, Doi 2007, Jalalian et al. 2013, Meiring 2015, Meyer 2001, Meyer 2004, Pritchett 2009, Robles et al. 2010, Vashitz et al. 2009].

The advantage of analyzing binary classification systems is that, in this case, the ResQu model is reduced to relatively simple calculations and interpretations [Douer & Meyer 2020]. This also makes it easier to control and examine the effects of different human and system characteristics on responsibility.

Let $X$ denote the binary set of the action alternatives for the human user, and $Y$ denote the binary classification set for the system. Then, the ResQu model defines human responsibility in this case as

$$Resp(X) \stackrel{\text{def}}{=} \frac{H(X/Y)}{H(X)} = \frac{H(X,Y) - H(X)}{H(X)} \tag{1}$$

where $H(X)$ is Shannon's entropy, which is a measure of uncertainty related to a discrete random variable $X$

$$H(X) = -\sum_{x \in \chi} p(x) log_2 p(x) \tag{2}$$

and $H(X/Y)$ is the conditional entropy, which is a measure of the remaining uncertainty about a variable $X$ when a variable $Y$ is known.

$$H(X/Y) = -\sum_{y \in Y} p(y) \sum_{x \in \chi} p(x/y) \, log_2 p(x/y) \tag{3}$$

The ratio $Resp(X)$ in (1) quantifies the expected exclusive share of the human in determining the human action selection $X$, given the system's classification $Y$. It is computed from the entropy $H(Y)$ of the system classification results, the entropy $H(X)$ of the human action selection, and their joint entropy $H(X,Y)$. By definition, $Resp(X) \in [0,1]$. $Resp(X) = 1$ if, and only if, the human action selection $X$ is independent from the system's classification result $Y$, in which case the human is fully responsible for the system output. $Resp(Z) = 0$ if, and only if, $Y$ completely determines $X$, in which case the human actions are exclusively determined by the system's classifications, so the human's comparative responsibility is zero.

## 2.3 Signal Detection Theory for Modeling Aided Decision Making

In the current study, we used the framework of signal detection theory (SDT) [Green and Swets 1966], and specifically the basic equal variance Gaussian SDT model, to compute ResQu model predictions of human responsibility in the interaction with binary classification systems (i.e. theoretical responsibility) and to calculate the actual responsibility participants assumed (i.e. measured responsibility).

In terms of SDT, aided decision making is the combined performance of two detectors, the human and the classification system [Maltz & Meyer 2001, Meyer 2001, Meyer & Ballas 1997, Sorkin 1988, Sorkin & Woods 1985]. Both detectors obtain some information on the state of the environment, which is probabilistically related to the actual state of the environment. The two sources are imperfectly correlated, because otherwise they would be redundant.

According to SDT, in binary classification systems, detectors observe an ambiguous stimulus and have to decide to which of two possible categories it belongs. It is customary to refer to the rare event that needs to be detected (a cyberattack, a malfunction, a pathology, a crime, etc.) as the signal. The prior probability that an observed stimulus is a signal or noise, is determined by their relative prevalence in the environment and will be denoted by $P_s$ and $1 - P_s$, respectively. Both types of stimulus can be measured by a single observable parameter, which transforms the data into a continuous scale value (which will be referred to as the continuous information). The distributions of the values differ for signal and noise entities (with signal usually assumed to have a higher mean value than noise), which allows some discrimination between the two types. However, the distributions overlap, so when a certain value is observed, there is ambiguity whether the stimulus is indeed a signal, or whether it is actually noise.

In describing the detector, SDT differentiates between its detection sensitivity and its response criterion. The detection sensitivity ($d'$) is the detector's ability to distinguish between signal and noise. In the basic equal variance Gaussian SDT model, the detection sensitivity is defined as the distance between the means of the signal and noise Gaussian distributions, and is measured in standard deviations. It can be represented as a shift of the signal probability density function, compared to the noise probability density function. When $d'=0$, the detector is unable to distinguish between signal and noise. The larger the detection sensitivity, the easier it is for the detector to distinguish between signals and noise. We will denote by $d'_A$ and $d'_H$ the detection sensitivities of the binary classification system and the human, respectively.

For every value of the observed parameter, it is possible to compute the likelihoods of observing the value under the signal distribution or the noise distributions. SDT assumes a threshold likelihood ratio, which is called the response criterion ($\beta$), or response bias. The value of the observable parameter at the threshold serves as a cutoff point for the detector. When the observed value is below the cutoff point, the detector classifies the observation as noise, and above the cutoff point, as a signal. The binary classification system has a preset response criterion, denoted by $\beta_A$, which is used to determine its output by comparing the acquired value to a preprogrammed fixed cutoff point. When the human works alone, without the use of a decision aid, the optimal human response criterion, $\beta_H$, that maximizes the expected value of the payoffs is given by:

$$\beta_H^* = \frac{1-P_S}{P_S} \frac{V_{TN}-V_{FP}}{V_{TP}-V_{FN}} \tag{4}$$

Where $V_{TN}$, $V_{FP}$, $V_{TP}$, and $V_{FN}$ represent the human utility values for True Negative (correctly responding "noise"), False Positive (falsely responding "signal"), True Positive (correctly responding "signal") and False Negative (falsely responding "noise"), respectively.

When aided by a binary classification system, the classification output of the system serves as additional input for the human, who can combine the information from the system with self-acquired information. The human can improve the decision-making process by judging the values of the continuous observable parameter with different response criteria, depending on the classification results of the system [Robinson & Sorkin 1985]. The different response criteria are computed by replacing $P_S$ in (4), with $P_{S|A}$ - the conditional probability for a signal, given that the system classified the stimulus as a signal ("Alarm"), or $P_{S|NA}$ - the conditional probability for a signal, given that the system classified the stimulus as noise ("No Alarm"). When using a reliable classification system, the posterior probability for a signal is larger when an alarm is issued, and it is smaller when there is no alarm, $P_{S|A} \geq P_S \geq P_{S|NA}$. In this case, the user should adopt a lower cutoff point when alarm is issued (i.e., increase the tendency to declare a signal) and a higher cutoff point when no alarm is issued (i.e., increase the tendency to declare noise when the system indicated that "all is well").

## 2.4 Signal Detection Measures for Trust, Compliance and Reliance in Aided Decision Making

The above theoretical SDT formulation assumed a best-case scenario of perfect rationality on the part of the human, perfect human knowledge of the system's properties and optimal human utilization of information. Under these assumptions, the computed human responsibility is optimal, given the properties of the system. However, in reality, people may act non-optimally with intelligent systems by misusing, disusing, and abusing the system [Alvarado-Valencia & Barrero 2014, Arnott 2006, Baker et al. 2004, Goddard et al. 2011, Mosier et al. 1998, Parasuraman & Riley 1997].

Human trust in automation, which reflects the human's attitude toward the system, is a major factor that influences how people use intelligent systems in real-life [Lee & See 2004]. Thus, it has become a central issue in the study of human interaction with intelligent systems [Lee & Moray 1994, Muir & Moray 1996, Meyer & Lee 2013, Meyer et al. 2014]. Human trust in systems is influenced by factors associated with the human (e.g., culture, age, gender, cognitive and emotional factors, self-efficacy, expertise, workload, etc.), factors associated with the system (e.g., reliability, types of errors, usefulness, feedback, design features, etc.), and factors associated with the environment (mainly the predictability of the environment) [Alvarado-Valencia & Barrero 2014, Hoff & Bashir 2015, Schaefer et al. 2016, Sutherland et al. 2015, Sutherland et al. 2016].

There are two different possible responses to binary classification systems, such as binary alert systems, termed as reliance and compliance [Meyer 2001]. Compliance describes the response when an alarm was issued (i.e. an indication for a signal), whether true or false. A compliant operator will rapidly switch attention from the current activity to the alarm and will initiate a response. Reliance refers to the operator state when the system is silent, signaling "all is well" (i.e. an indication for noise). Reliant operators will allocate resources to other tasks, relying on the automation to let them know when a problem occurs [Dixon & Wickens 2006]. Compliance and reliance are somewhat independent responses that are affected by different factors [Meyer 2004].

In previous studies of human trust in systems, the human perception and behavior have been assessed through subjective ratings of trust, together with measuring the proportion of times during which the system was used [e.g., Muir & Moray 1996] or by assessing the probability of detecting system failures [e.g. Mosier et al. 1998]. However, in the interaction with binary classification systems, one can use the SDT-based measures as an alternative method for quantifying human trust, compliance and reliance [Meyer 2001, Meyer & Lee 2013].

The human's differential adjustment of the cutoff points to the output of the classification system is directly related to the level of human trust and causal human responsibility. When the human uses a single cutoff point, regardless of indications from the classification system, he or she obviously ignores the system's indications and has no trust in the system. The use of different cutoff points indicates that the human considers the information from the classification system when making a decision. The larger the difference between the cutoffs, the greater the trust and the weight the human gives to the information from the system. The levels of human reliance and compliance can be assessed, respectively, by the distance of the human cutoff point from zero and from the optimal corresponding value, when the classification system indicated signal or noise (i.e., an alarm or "all is well").

Another SDT measure related to human trust is the *effective d'*, denoted by $d'_{eff}$, which reflects the overall detection sensitivity, based on the combined detection capabilities of the system and the human. For a binary classification system, the maximal value of $d'_{eff}$ the human can attain can be approximated by [Meyer & Kuchar 2019]:

$$d'_{eff} = \sqrt{d'^2_H + d'^2_A - 0.3 d'^2_H d'^2_A} \qquad (5)$$

When the human under-trusts the classification system, the human will use information from the system in a non-optimal way, giving less weight to the system's indications. In this case, the empirical value of $d'_{eff}$ will be lower, and closer to the human's own detection sensitivity, $d'_H$, than to the maximal optimal value.

To conclude, differences between the optimal theoretical values and the empirical values of the cutoff differences and the effective $d'$ indicate non-optimal user trust in the classification system, which may lead to non-optimal user behaviors, such as over- or under-reliance and compliance. In the current study we used both SDT measures (i.e. the cutoff differences and the effective $d'$) to investigate the underlying causes for differences between theoretical and measured responsibility.

## 3 EXPERIMENT 1: THE EFFECT OF $d'$ ON HUMAN RESPONSIBILITY

### 3.1 Introduction and Objectives

According to the ResQu model, when the human and an intelligent system have similar incentives, the relative abilities of the human and the intelligent system are the main determinants of the human's responsibility. When the system ability exceeds that of the human, the human's unique contribution diminishes, leading to low human causal responsibility. Conversely, when the human's ability exceeds that of the system, the human's causal responsibility will be high. When aided by a classification system, the relative abilities of the human and the classification system are defined by their detection sensitivities.

Experiment 1 examined the ability of the ResQu model to predict human behavior and perceptions of responsibility for different combinations of the human's and system's detection sensitivities ($d'_H$ and $d'_A$, respectively), when both the human and the classification system have similar incentives regarding the outcomes (i.e. use a similar response criterion).

The objectives of the experiment were:

(1) To examine whether $d'_A$, $d'_H$, or their interaction, have significant effects on the measured and subjective responsibility, as predicted by the ResQu model (against the null hypothesis that they have no effect), and specifically that human responsibility decreases in $d'_A$, and increases in $d'_H$.

(2) To examine how close the values of the empirical responsibility are to the theoretical values and to analyze sources for differences by using common SDT measures of trust (difference between cutoffs and effective $d'$).

(3) To examine the relations between measured and subjective responsibilities.

(4) To examine whether the subjective responsibility estimates differ between self and another agent (against the null hypothesis that they do not).

## 3.2 Method

### 3.2.1 Selection of Experimental points

We examined different combinations of human detection sensitivity $d'_H$ and system detection sensitivity $d'_A$, each on a scale ranging between .6 (very low ability to distinguish between signal and noise) and 3 (high ability to distinguish between signal and noise). For each pair of human and system detection sensitivities, we calculated a predicted theoretical ResQu responsibility value [Douer & Meyer 2020], using signal probability and payoff scheme values, which are described below in the procedure and design section. The calculation assumed that both the human and the binary classification system have similar incentives regarding the outcomes, and thus they use the same response criterion. Fig. 2 depicts the computed ResQu responsibility values as a function of $d'_H$ and $d'_A$.

We then selected four experimental points in which the human and the classification system either have a low detection sensitivity ($d'_H$, $d'_A$ = 1) or high detection sensitivity ($d'_H$, $d'_A$ = 2.3). These points have the advantage of spanning the theoretical responsibility range, from as low as 12% to as high as 87% (see Fig. 2).

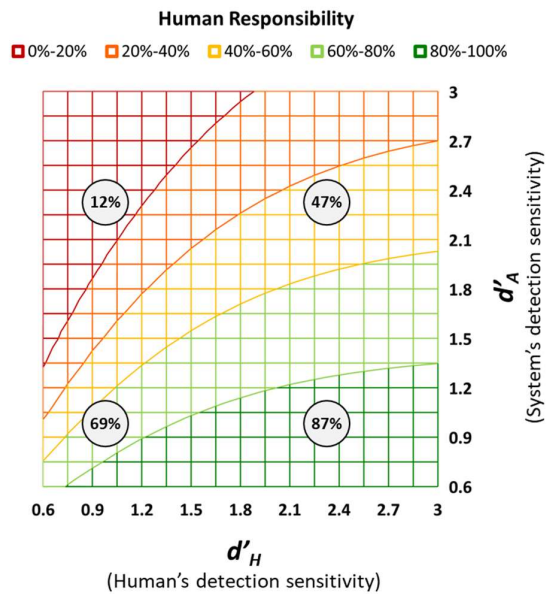

Fig. 2. ResQu model theoretical responsibility values for different combinations of human and system detection sensitivities, ranging from .6 (very low ability to distinguish between signal and noise) to 3 (very high ability to distinguish between signal and noise). The circles present the four selected experimental points, and their computed responsibility values.

### 3.2.2. Participants

Participants were 60 students from Tel Aviv University (ages 20-49, median 24, 62% females), of which 53 were undergraduate students and 48 belonged to the Faculty of Engineering. They were recruited through E-mail invitations and a post on a university webpage that serves to recruit students for experiments. We assigned the participants randomly to one of the four experimental subgroups, so that we had 15 participants in each subgroup. Each participant received 40 Israeli New Shekels [ILS], about US $12, for taking part in the experiment. Conscientious performance of the task was encouraged by the promise of an additional monetary award (100 ILS, about US $29) to a randomly selected participant in each of the experimental groups, using the accumulated individual scores as weights for the selection.

### 3.2.3. Apparatus

The experiment was conducted on desktop computers, with Intel® i7 3.4 GHZ Processor, 8 GB RAM, NVIDIA® GeForce GT 610 Video Card, and 23-inch (56-cm) monitors. The experimental program was written in Python.

Fig. 3 shows a schematic depiction of the experimental screen. It consisted of a 20 cm high and wide square at the center of the screen. Above the square were two fields, labeled "Total Score" and "Last Trial", which displayed the cumulative number of points and the number of points that were gained or lost in the last trial.

The participants observed an ambiguous continuous stimulus in the form of a rectangle, displayed for 30 seconds inside the large square. The rectangle had a fixed width, but its height varied between trials, as it was sampled from one of two distributions of either long or short rectangles. In each trial, participants had to decide whether the rectangle was from the long or the short distribution. By defining the level of overlap between the distributions, we could control and assign different levels of human detection sensitivities ($d'_H$) to different experimental groups. Similar to a method used in previous studies [Meyer 2001], in each trial the rectangle appeared at a different position inside the large square to make it more difficult for participants to mark the cutoff point explicitly by, for instance, placing their finger on the screen.

The experimental binary stimulus from the classification system was represented by a small square, located at the top of the screen, that could have one of two possible colors, either red (F44141 Hex color code), indicating system classification as signal, or green (4EF442 Hex color code), indicating noise. The binary stimulus appeared together with the rectangle stimulus, and it remained visible while the rectangle was shown.

Participants responded by clicking with the mouse on either the "Accept" or the "Reject" button at the bottom of the screen, according to whether they thought that the rectangle belonged to the longer or the shorter distribution. After the response, the payoff for the trial appeared in the "Last Trial" field, and the "Score" field was updated. An additional feedback message, stating either "correct" or "incorrect", appeared for 2 seconds, and then the next trial began.

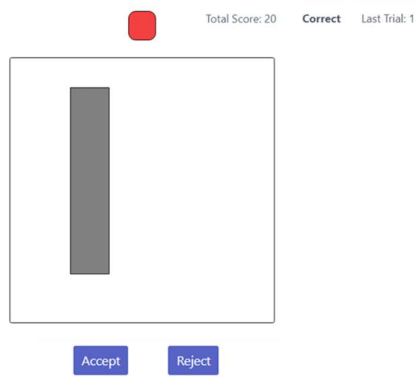

Fig. 3. A schematic depiction of the experimental screen. The figure presents an example in which there is an indication for signal (the indicator field is red), the cumulative number of points is 20, and the participant chose a correct response in the last trial, which awarded an additional point.

## 3.3. Procedure and Design

The experiment was conducted in the "Interaction with Technology (IwiT) Lab" of the Industrial Engineering department at Tel Aviv University on groups of up to 7 participants. Each participant sat at a computer.

The instructions stated that the experiment is a simplified simulation of a quality control task in a factory. A certain percentage of the items the factory produces are defective. A quality control worker inspects and classifies each produced item and decides if it is "intact" and should be accepted, or a "defect" that should be rejected. The worker makes a decision, based on the vertical length of the rectangles. Intact items have a shorter mean length than defective items, but the two distributions overlap. Thus, when the worker observes the length of a rectangle, uncertainty remains if the item is intact or defective. Participants were told that the factory considers acquiring a system that will aid the worker in the classification task. The system classifies each item independently. The classification results appear as either a red indication, when the system identifies a potentially defective item, or a green indication, when the system identifies an item as intact. The factory considers two optional classification systems, which may differ in their classification accuracy. Participants were told that their mission is to rate and compare the performance and contribution of the two candidate systems.

The theoretical ResQu model assumes a fixed human detection sensitivity over time. As such, it does not consider changes in the detection sensitivity over time, due to learning effects or fatigue. We dealt with this issue when planning the experiment by enabling a learning period and restricting the overall number of trials. Each participant performed 100 trials with each of the two systems, deciding on each trial whether or not to reject or to accept an item, based on the visual inspection of the item's length and the binary indication given by the classification system. The 100 trials with each system were divided into two blocks, each with 50 trials. The participants were told that the first block of 50 trials was mainly for learning and gaining experience with their own abilities and the abilities of the system, and that their performance will be assessed according to their score in the second block. For a very simple task and simplified settings, such as those used in the experiment, the first block of 50 trials is sufficient to exhaust the learning effect, while the second block of 50 trials is short enough to avoid fatigue effects.

In each block of 50 trials, for 30 trials the stimuli were sampled from the short distribution (intact items), and for 20 trials they were sampled from the long distribution (defective items), representing a probability of .4 for a defective item. The trials were individually randomized for each participant, system response and block.

Participants received 1 point for any correct rejection of a defective item (True Positive) or acceptance of an intact item (True Negatives). Participants lost 1 point for rejecting an intact item (False Positive) and lost 2 points for accepting a defective item (False Negative). This payoff scheme reflects a factory's incentive not to deliver defective items to costumers, which is stronger than the incentive not to reject intact items. When a human conducts the classification task without the aid of the classification system, the optimal human response criterion for the above payoff scheme and .4 probability for a defective item, is $\beta_H = 1$ (see equation 4.)

The 60 participants were randomly assigned to one of two equal size groups, which differed in the level of overlap between the two distributions of long and short rectangle population, which was set to represent the two selected levels of human detection sensitivity of either $d'_H = 1$ ("less-accurate" human) or $d'_H = 2.3$ ("accurate" human). All participants saw classification results from both a "less accurate" system that had a detection sensitivity of $d'_A = 1.0$ and from an "accurate" system that had a detection sensitivity of $d'_A = 2.3$, in two parts of the experiment. Both classification systems used a response criterion of $\beta_A = 1$, which matches a participants' optimal response criterion. The order of experiencing the systems was counterbalanced so that, in each group, half of the participants were aided by the accurate classification system in the first part and by the less-accurate classification system in the second part, and the other half of the participants saw the systems in reversed order. Thus, there were four experimental subgroups in the experiment, created by combinations of two levels of the participant's detection sensitivity ($d'_H = 1$ and $d'_H = 2.3$) and the order in which the participants examined the two types of classification systems ($d'_A = 1$ first or $d'_A = 2.3$ first). Table 1 summarizes the four experimental subgroups.

TABLE 1. The four experimental subgroups in Experiment 1

| Human detection sensitivity | $d'_H = 1$ | | $d'_H = 2.3$ | |
|---|---|---|---|---|
| Order of examining | $d'_A$=1 | $d'_A$=2.3 | $d'_A$=1 | $d'_A$=2.3 |
| the classification Systems | $d'_A$=2.3 | $d'_A$=1 | $d'_A$=2.3 | $d'_A$=1 |
| # of participants in each subgroup | 15 | 15 | 15 | 15 |

Table 2 summarizes the outcome probabilities for the two classification systems, used in Experiment 1, and presents their positive and negative predictive values (PPV and NPV, respectively), which are, respectively, the probabilities for a defective item when the system classified the stimulus as a defective item (i.e. presented a red indication) and for an intact item when the system classified the stimulus as an intact item (i.e. presented a green indication).

TABLE 2. Outcome probabilities for the two classification systems in Experiment 1

| Type of System | Parameters | Defect (Signal) | | Intact (Noise) | | PPV | NPV |
|---|---|---|---|---|---|---|---|
| | | Red True Positive | Green False Negative | Red False Positive | Green True Negative | | |
| **Less-Accurate** | $d'_A$=1.0, $\beta$=1 | 69% | 31% | 31% | 69% | 60% | 77% |
| **Accurate** | $d'_A$=2.3, $\beta$=1 | 87% | 13% | 13% | 87% | 82% | 91% |

After completing the trials with each system, participants filled out a questionnaire, providing their subjective judgments on the accuracy of the system and its contribution to their performance. In each question, the participants rated their level of agreement on a scale between 1 (not at all) and 7 (very much). Table 3 presents the questions and the factors to which they relate.

TABLE 3. Questions for the subjective assessment of the system sensitivity, human sensitivity and human responsibility

| Factor | Question # | Question |
|---|---|---|
| **System detection sensitivity- $d'_A$** | Q1 | The classification system could distinguish between intact and faulty items |
| **Human detection sensitivity- $d'_H$** | Q2 | I could distinguish by myself (without the aid of the system) between intact and faulty items |
| **Self Responsibility (Contribution to action selection)** | Q3 | I used the indications from the classification system to select an action |
| | Q4 | When selecting an action, I relied more on the indications from the classification system than on my own detection abilities |
| | Q5 | The classification system had a low contribution - I could have similar performance without it |
| **Responsibility of another agent** | Q6 | When aided by this system, a human remains responsible for the classification process |

## 3.4. Results and Discussion

### 3.4.1. Measured Responsibility

According to the ResQu model, theoretical responsibility monotonically decreases in $d'_A$ and monotonically increases in $d'_H$. We analyzed the measured responsibility (i.e. the actual observed human contribution) with a three-way mixed analysis of variance (ANOVA), with human detection sensitivity and the order in which the classification systems were examined as between-subjects variables and the type of classification system as a within-subjects variable. There was no significant main effect of the order of experiencing the two classification systems, nor any significant interaction that involved the order. Thus, we focus on the results for the remaining two variables and their interaction. Table 4 summarizes the results. For effect size we report both partial eta squared ($\eta^2$) and generalized eta squared ($\eta^2_G$), which are the most appropriate measures for mixed design studies that include both between-subjects and within-subjects effects (Bakeman 2005, Olejnik and Algina, 2003)

TABLE 4. ANOVA results for measured responsibility

| Variable | Effect on Measured Responsibility | | | | |
|---|---|---|---|---|---|
| | F(1,56) | MSE | Par. $\eta^2$ | $\eta^2_G$ | Observed power |
| $d'_A$ | 122.84 **** | 2.32 | .69 | .43 | 1 |
| $d'_H$ | 9.60 *** | .35 | .15 | .10 | 0.86 |
| $d'_A X d'_H$ | 1.61 | .03 | .03 | .01 | 0.24 |

\* p < .05; \*\* p < .01; \*\*\* p < .005; \*\*\*\* p < .0001

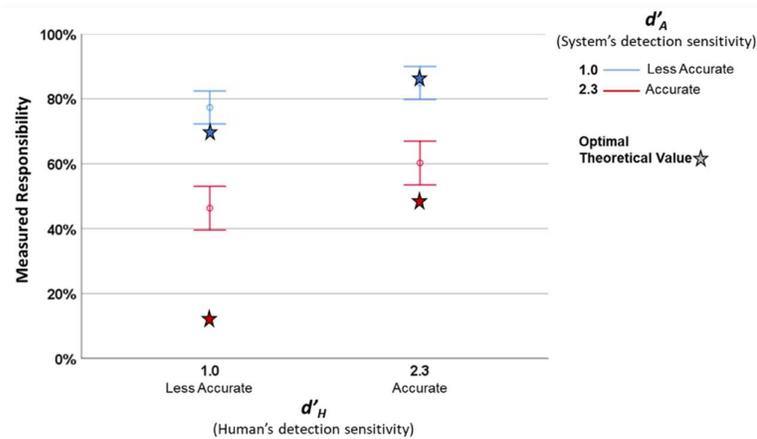

Fig. 4. Mean values for measured responsibility. Error bars represent 95% CI. Star icons indicate the optimal theoretical values from the ResQu model.

Fig. 4 depicts the mean values of empirical measured responsibility and the model predictions of the optimal theoretical values. As predicted by the ResQu model, both the human's and the system's detection sensitivities had significant effects on the measured responsibility, with a large effect for the system detection sensitivity and a smaller effect for the human detection sensitivity. The measured responsibility indeed decreased in $d'_A$ and increased in $d'_H$. All participants relied less on information from the less accurate classification system, leading to higher measured human responsibility with this system. In addition, the accurate participants tended to rely less on each of the systems than the less accurate participants, so their measured responsibility with each system was higher.

Fig. 4 shows that in most cases, the mean value of the measured responsibility was close to the optimal predicted theoretical value, except for the case in which the less accurate participants used the accurate classification system. In this case, the less accurate participants assumed much higher-than-optimal responsibility.

A difference between the measured responsibility and the optimal theoretical value can be due to participants' inadequate use of the information they have independently or to participants giving excessive or too little weight to the information from the classification system. It is possible to distinguish between these two sources by analyzing the SDT measures - the effective sensitivity $d'_{eff}$ and the difference between cutoffs points. Fig. 5 presents the optimal and mean empirical values for $d'_{eff}$ and the cutoff differences. Table 5 summarizes the empirical deviations from theoretical predictions for both the ResQu model and SDT.

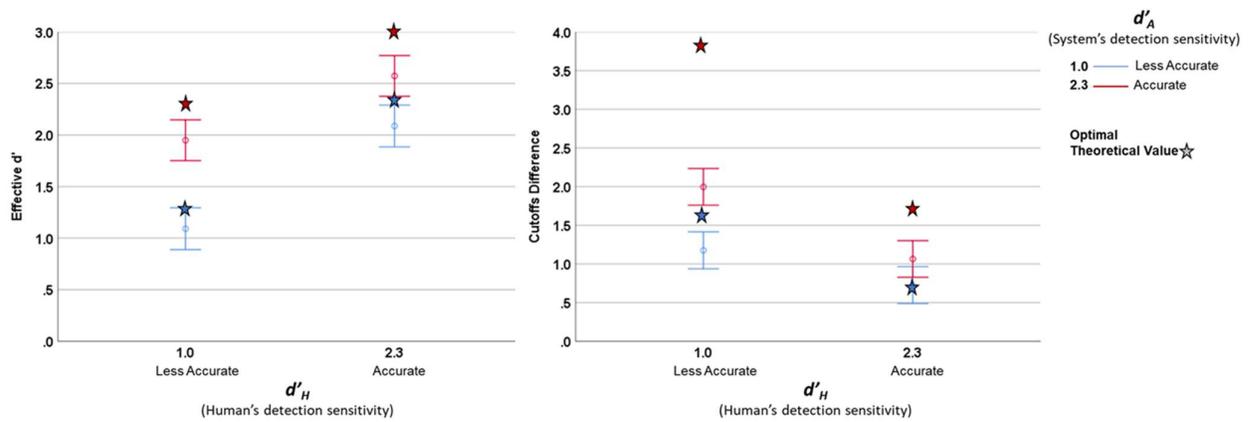

Fig. 5. Mean empirical values for $d'_{eff}$, and cutoff differences. Error bars represent 95% CI. The star icons indicate the optimal theoretical values of SDT.

TABLE 5. Theoretical predictions vs. empirical results

| Human | System | ResQu - Measured Responsibility | | | SDT - $d'_{eff}$ | | | SDT - Cutoffs Difference | | |
|---|---|---|---|---|---|---|---|---|---|---|
| $d'_H$ | $d'_A$ | Theoretical Optimum | Empiric Mean | Differ. | Theoretical Optimum | Empiric Mean | Differ. | Theoretical Optimum | Empiric Mean | Differ. |
| **1** | **1** | 69% | 77% | 8% | 1.3 | 1.1 | -0.2 | 1.6 | 1.2 | -0.4 |
| | **2.3** | 12% | 46% | 34% | 2.3 | 2.0 | -0.3 | 3.9 | 2.0 | -1.9 |
| **2.3** | **1** | 87% | 85% | -2% | 2.3 | 2.1 | -0.2 | 0.7 | 0.7 | 0 |
| | **2.3** | 47% | 60% | 13% | 3.0 | 2.6 | -0.4 | 1.7 | 1.1 | -0.6 |

In all combinations of human and system detection sensitivities, the empirical $d'_{eff}$ was lower than the corresponding optimal theoretical value by 10%-15%. Thus, participants did not optimally combine their own information with that of the classification system. In particular, with the accurate system, less accurate participants reached $d'_{eff} = 2.0$, which was lower than the system sensitivity (2.3). They would have fared better if they would have entirely ignored the continuous stimulus and responded only to the indications from the classification system.

In most combinations of human and system detection sensitivities, the cutoff difference was lower than the optimal value, implying that participants tended to under-trust the indications from the classification systems. The deviations from the optimal value were larger when participants used the accurate system and were the largest when the less accurate participants used the accurate system. This deviation from the optimal value parallels that of the measured responsibility.

To conclude, an analysis of traditional SDT measures of trust suggests that the deviation of the measured responsibility from the optimal theoretical value is mainly due to less accurate participants overestimating their own capabilities, compared to those of the accurate classification system. This led them to select non-optimal cutoff points and to use the information from the system in a non-optimal way. This result is in line with previous results from behavioral research in SDT [Bartlett & McCarley 2017, Maltz & Meyer 2001, Meyer 2001, Meyer et al, 2014], in which users tended to overestimate their own capabilities, especially when they performed poorly.

### 3.4.2. Subjective self-responsibility

Question *Q1* referred to the subjective assessment of the system's classification ability and question *Q2* referred to the subjective assessment of the participant's classification ability. Questions *Q3-Q5* referred to the participants' subjective assessments of their own responsibility. We reverse-scored questions *Q3* and *Q4* for the score to reflect responsibility and performed a reliability analysis to measure the consistency of the questions. The analysis showed high reliability, with Cronbach's $\alpha$ = .87, so we used their average as an estimate for the subjective responsibility.

We analyzed the different subjective assessments with three-way mixed analyses of variance (ANOVA), with human sensitivity and the order of the classification systems as between-subject variables, and the type of the classification system as a within-subjects variable.

The only significant effect involving the order of experiencing the two classification systems was the three-way interaction between order, human sensitivity and system sensitivity in the subjective assessments of the system's classification ability (question *Q1*), which had a small effect, $F_{(1,56)}= 6.81, MSE=8.53, Par. \eta^2= .11, \eta^2_G = .05, p =.01$. In this case, the order only had a significant effect on the accurate participants' ratings of the performance of the less-accurate classification system. These participants assessed the performance of the less-accurate system to be better (average score of 5.2) when it was examined first than when it was examined second (average score of 3.1). The order had no significant effect on how the accurate participants rated the performance of the accurate classification system or how the less-accurate participants rated both systems. In all other questions, there was no significant main effect of the order, nor any significant interaction that involved the order. Thus, we focus on the results for the remaining two variables and their interactions. Table 6 summarizes the ANOVA results.

In the analysis of question *Q1*, the subjective assessment of the systems' classification abilities, the only additional significant factor, with a large effect size was the actual difference between the two classification systems. The accurate system was rated as having significantly higher ability (*mean*= 5.4, *Sd*= .14) than the less accurate system (*mean*= 3.9, *Sd*= .16). Thus, irrespective of their own abilities, both types of participants noticed that one system had a higher ability.

In the analysis of question *Q2*, the subjective assessment of participants' own classification ability, only the actual difference between the detection sensitivities of the two participants groups was significant, with a medium effect size. The more accurate participants rated their classification abilities significantly higher (*mean*= 5.3, *Sd*= .14) than the less accurate participants (*mean*= 4.5, *Sd*= .14). Thus, participants were clearly able to evaluate their own performance, independently from the performance of the classification systems.

The mean score for (reverse coded) questions *Q3, Q4* and question *Q5* reflected the participants' subjective assessments of their own responsibility. In this case, both the system and the human sensitivities were very significant, with a large and a medium effect size, respectively. Their interaction was also significant, but to a much lesser degree and had only a small effect size. The accurate participants rated their responsibility with the accurate system significantly lower (*mean*= 3.9, *Sd*= .21) than with the less accurate system (*mean*= 4.9, *Sd*= .23). The less accurate participants also rated their responsibility with the accurate system significantly lower (*mean*= 2.5, *Sd*= .21) than with the less accurate system (*mean*= 4.3, *Sd*= .23). The interaction between the participants and the system types was due to the less accurate participants differentiating more between the two types of classification systems.

TABLE 6. ANOVA results for questions *Q1*, *Q2*, *Q3-Q5*

| *Variable* | **Q1- Subjective assessment of system's classification ability** | | | | | **Q2- Subjective assessment of participant's classification ability** | | | | | **Q3-Q5 - Subjective assessment of participant's responsibility** | | | | |
|---|---|---|---|---|---|---|---|---|---|---|---|---|---|---|---|
| | F(1,56) | MSE | Par. η2 | $\eta^2_G$ | Obs. Power | F(1,56) | MSE | Par. η2 | $\eta^2_G$ | Obs. Power | F(1,56) | MSE | Par. η2 | $\eta^2_G$ | Obs. Power |
| $d'_A$ | 51.48 **** | 64.53 | .48 | .29 | 1.00 | 1.72 | 1.87 | .03 | .01 | .25 | 51.36 **** | 63.07 | .48 | .27 | 1.00 |
| $d'_H$ | 2.19 | 3.33 | .04 | .02 | .31 | 14.72 **** | 18.41 | .21 | .12 | .96 | 16.64 **** | 29.67 | .23 | .15 | .98 |
| $d'_A X d'_H$ | 2.15 | 2.7 | .04 | .02 | .30 | .62 | .68 | .01 | .01 | .12 | 4.71* | 5.78 | .08 | .03 | .58 |

*\* p < .05; \*\* p < .01 ; \*\*\*p < .005 ; \*\*\*\* p < .0001*

To conclude, in line with the theoretical predictions of the ResQu model, the subjective assessment of responsibility decreased in $d'_A$ and increased in $d'_H$. In addition, the subjective assessment of responsibility was coherent with the subjective assessments made in *Q1* and *Q2* regarding the system's and the human's classification abilites.

The pattern of subjective responsibility scores resembles that of the measured responsibility. To compare the two empirical results, we normalized the mean scores of the two responsibilities. The normalized values of mean subjective responsibility were close to those of the measured responsibility (see Fig. 6). This implies that participants judged their marginal contribution with each classification system quite accurately.

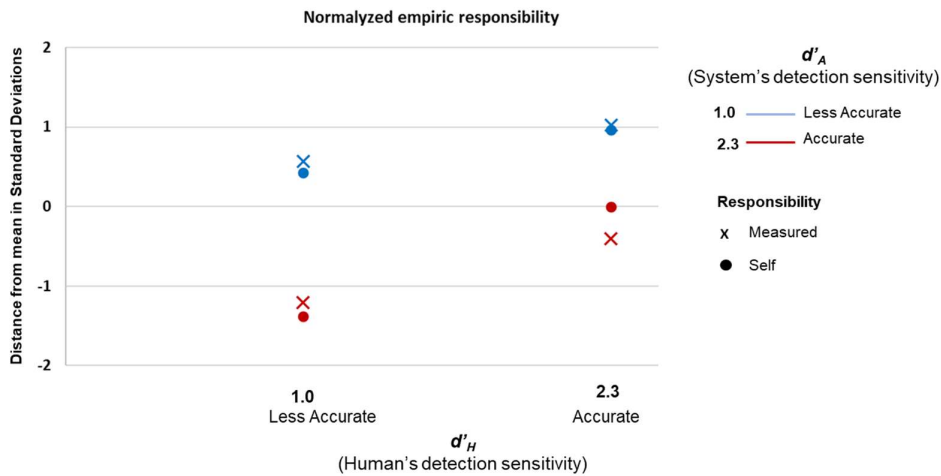

*Fig.* 6. Normalized measured responsibility Vs. normalized subjective responsibility (based on *Q3-Q5*)

### 3.4.3. Measured responsibility and subjective assessments as mediators

In Experiment 1, we used a manipulated proxy for theoretical responsibility, in the form of different combinations of human and system detection sensitivities. We showed a significant effect on the corresponding measured and the subjective responsibilities. The ANOVA results revealed that, as predicted by the ResQu model, both the measured and the subjective responsibilities were decreasing in $d'_A$ and increasing in $d'_H$.

The subjective assessments were provided by filling a questionnaire, after completing the trials with each classification system. Thus, one may hypothesize that the actual "hands on" experience with each system (i.e. the measured responsibility), may have served as a mediator in the forming of subsequent the subjective responsibility assessments. To test this mediation hypothesis, we used an analytical approach outlined by Preacher and Hayes (2004) and Shrout and Bolger (2002). It directly tests the indirect effect between a predictor and an outcome variable through one or more mediators via a bootstrapping procedure. We first analyzed a simple mediation model, which examined whether the direct effect of the manipulated proxy for theoretical responsibility on the subjective perception of responsibility (expressed in questions *Q3-Q5*), was partially or fully mediated by the measured responsibility. Fig. 7 summarizes the model's results. All the coefficients in the figure are fully standardized and the confidence intervals are bias corrected bootstrapped CI based on 5000 samples.

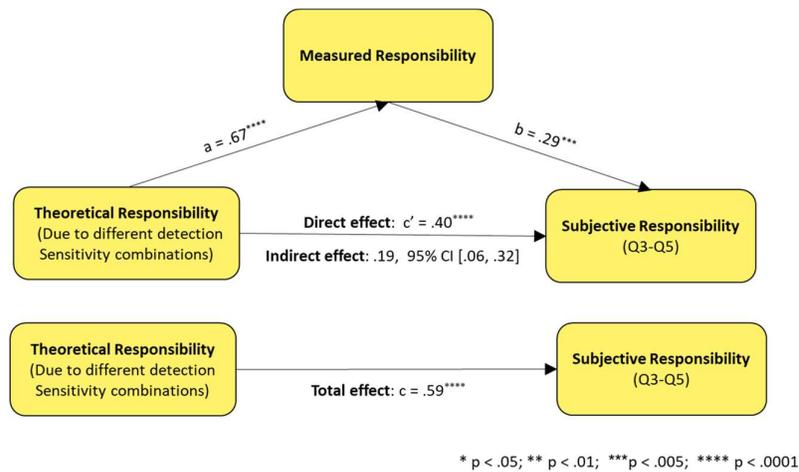

*Fig.* 7. Model of theoretical responsibility as a predictor of subjective responsibility, mediated by measured responsibility.

Coherent with the ANOVA results in Table 4 and Table 6, Fig. 7 presents a significant effect of theoretical responsibility on both the measured and the subjective responsibilities. More importantly, the results show a significant indirect effect of theoretical responsibility on subjective responsibility via partial mediation by measured responsibility, *b*=0.19 (SE=.07), 95% BCa CI [.06, .32]. About 32% of the total effect of theoretical responsibility on subjective responsibility may be attributed to the indirect effect through measured responsibility. Hence, in an interesting manner, the formation of subjective responsibility was not based mainly on the actual level of human contribution, but probably on other subjective factors. Such factors could be the separate preceding subjective assessments of system capabilities (*Q1*) and human capabilities (*Q2*).

To examine this hypothesis, we expanded the mediation model to include parallel multiple mediation of these additional factors. Fig. 8 summarizes the results. All the coefficients in the figure are fully standardized, and the confidence intervals are bias corrected bootstrapped CI based on 5000 samples.

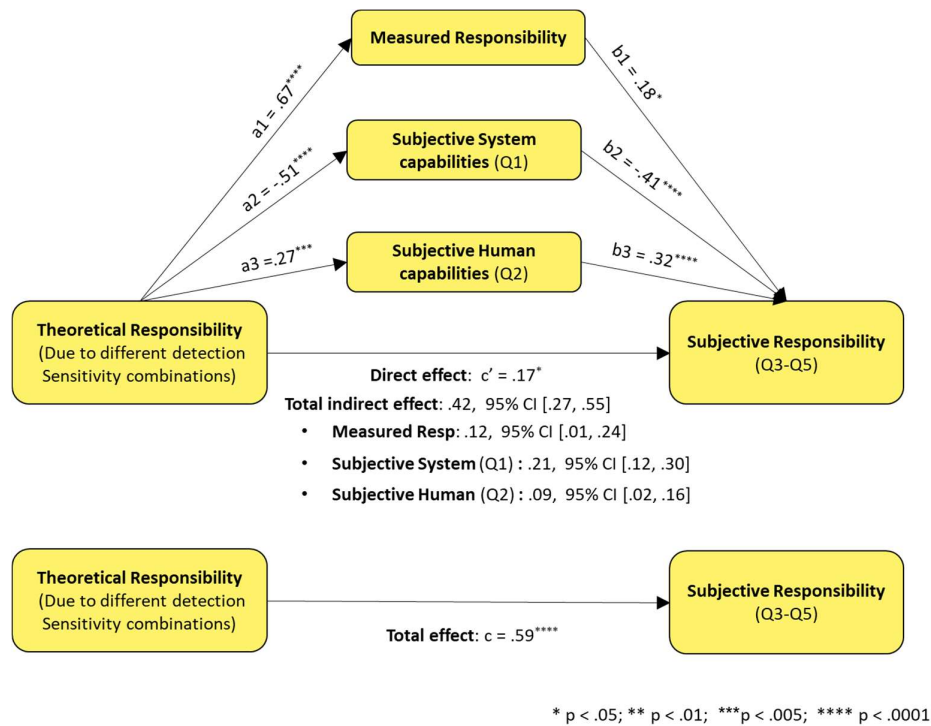

*Fig.* 8 A model of theoretical responsibility as a predictor of subjective responsibility, mediated by measured responsibility and subjective assessments of system and human capabilities.

The results show a large significant total indirect effect of theoretical responsibility on subjective responsibility through a mediation by both measured responsibility and the subjective assessments of system and human capabilities, $b$=.42 (SE=.07), 95% BCa CI [.27,.55]. After inclusion of the additional mediators, the total indirect effect constituted most of the total effect (71%), while the direct effect was reduced to a small portion (29%). Half of the total indirect effect is due to large and significant mediation via the subjective assessment regarding the system capabilities $b$=.21 (SE=.04), 95% BCa CI [.12,.30]. The rest of the total indirect effect was almost equally divided between significant mediations via subjective assessment regarding the human capabilities $b$=.09 (SE=.04), 95% BCa CI [.02, .16] and the measured responsibility $b$=.12 (SE=.05), 95% BCa CI [.01,.23].

Thus, there exists a significant large indirect effect of theoretical responsibility on subjective responsibility via mediation. As one may expect, the significant mediators were the subjective assessments of system and human capabilities, which together can be combined to form a subjective assessment of the relative human capabilities in respect those of the system, and the actual level of human unique contribution with the system (i.e. measured responsibility). Surprisingly, the most influencing mediator was the subjective assessment of system capabilities (i.e. how good the system was perceived). The measured responsibility and the perception of self-abilities, despite being also significant mediators, influenced to a lesser degree.

### 3.4.4. Subjective self-responsibility vs. another person's responsibility

Question *Q6* referred to the subjective assessment of another person's responsibility. We analyzed the different subjective assessments with three-way mixed analyses of variance (ANOVA), with human sensitivity and the order of the classification systems as between-subject variables, and the type of the classification system as a within-subjects variable. There was no significant main effect of the order, nor any significant interaction that involved the order. Thus, we focus on the results for the remaining two variables and their interactions. Table 7 summarizes the results.

TABLE 7. ANOVA results for Question *Q6*

| Variable | Q6 - Subjective assessment of another person's responsibility | | | | |
|---|---|---|---|---|---|
| | F(1,56) | MSE | Par. η2 | $\eta^2_G$ | Obs. Power |
| $d'_A$ | 33.86**** | 48.13 | .38 | .22 | 1.00 |
| $d'_H$ | 2.91 | 4.8 | .05 | .03 | .39 |
| $d'_A \times d'_H$ | 1.90 | 2.7 | .03 | .02 | .27 |

* $p < .05$; ** $p < .01$; *** $p < .005$; **** $p < .0001$

Only the differences between the classification systems were significant, with a large effect size. The two types of participants rated the responsibility of another person with the less-accurate system similarly (*mean*= 5.9, *Sd*= .17). This was significantly higher than their ratings of another person's responsibility with the accurate system, (*mean*= 5.0, *Sd*= .27 for the accurate participants and *mean*= 4.3, *Sd*= .27 for the less accurate participants).

The pattern of subjective assessment of another person's responsibility resembles that of the subjective self-responsibility. To compare between the two types of subjective responsibilities, we normalized their average scores, since there is no guarantee that they are judged the same by end users despite using similar scales. The normalized means are very close to each other (see Fig. 9), which implies that participants rated the responsibility of another person, in the same situation, similar to their own responsibility.

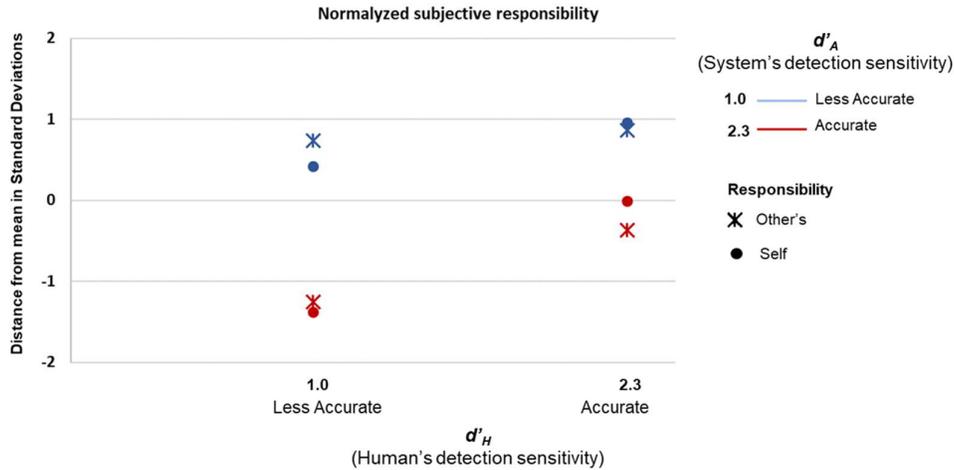

*Fig.* 9. Normalized subjective self-responsibility (based on *Q3-Q5*) and other's responsibility (based on *Q6*).

## 3.5. Conclusions for Experiment 1

Experiment 1 showed that, when the human and decision support systems have the same incentives, the ResQu model can serve as a descriptive model for both actual behavior (measured responsibility) and perceptions (subjective responsibility) at different combinations of the human's and the system's detection sensitivities ($d'_H$ and $d'_A$, respectively).

Regarding the objectives of the experiment, the results showed that:

(1) As predicted by the ResQu model, the classification system's detection sensitivity ($d'_A$) and the human's detection sensitivity ($d'_H$), had significant effects on both measured and subjective responsibilities, which were decreasing in $d'_A$ and increasing in $d'_H$.

(2) In most cases, the mean value of the measured responsibility was close to the optimal predicted theoretical responsibility value, except for the case in which the less accurate participants used the accurate classification system. In this case, the less accurate participants assumed much higher-than-optimal responsibility. An analysis, based on SDT measures of trust, showed that these participants overestimated their own capabilities, compared to those of the accurate classification system, which led them to select non-optimal thresholds.

(3) A comparison between measured and subjective responsibilities showed that participants subjectively judged their actual contribution with each classification system quite accurately. There was a significant effect of theoretical responsibility (represented by different combinations of detection sensitivities) on both measured and subjective responsibilities. However, the effect on subjective responsibility was largely indirect, via mediation. The most influencing mediator was the subjective assessment of system capabilities. Other significant mediators, but to a lesser extent, were the measured responsibility (i.e. the actual level of human contribution) and the subjective perception of self-abilities.

(4) Participants rated the responsibility of another person, in the same situation, similar to their own responsibility.

# 4 EXPERIMENT 2: THE EFFECT OF $\beta$ ON HUMAN RESPONSIBILITY

## 4.1 Introduction and Objectives

In Experiment 1, both the participants' and system's design were based on similar incentives, reflected by the use of a similar response criterion. However, designers of decision support systems may have perspectives and incentives that differ from those of the users, leading them to implement a different response criterion from the one the users adopt. Such use of different decision thresholds reflects a situation in which the human and the system designers have different estimates of the costs and benefits associated with different outcomes or different estimates of the signal and noise likelihoods.

For binary classification systems, the ResQu model [Douer and Meyer 2020] predicts that when the human's detection sensitivity is higher than that of the system, the human should rely mostly on own abilities, so differences in the system's response criterion have only a minor or no effect on human responsibility. Conversely, when the human's detection sensitivity is poor and inferior to the high detection sensitivity of the system, differences between the system's and the human's response criteria are predicted to have a major effect on human responsibility. In this case, the human should rely much less on a system which uses a considerably different response criterion and assume much higher responsibility, despite the system's far better detection abilities. This is a non-intuitive prediction of the theoretical model.

The scope of Experiment 2 was limited. It focused only on this single, but very interesting, case in which the human is predicted to assume high responsibility to compensate for the incentives difference, despite having poor detection abilities, which are inferior to those of the system. The objectives of Experiment 2 were as follows.

For the case in which the human has poor detection sensitivity and the classification system has good detection sensitivity:

(1) Examine whether, as predicted by the ResQu model, the human will assume much higher measured and subjective responsibility when the classification system uses a considerably different response criterion than the human's (against the null hypothesis that the difference in human's and system's response criteria have no effect).

(2) Examine how close the actual values of the empirical responsibility are to the theoretical values and to analyze sources for differences by using SDT measures of trust (effective $d'$ and difference between cutoffs).

(3) Examine the relations between measured and subjective responsibilities.

(4) Examine whether the subjective responsibility estimates differ between self and another agent (against the null hypothesis that they do not).

## 4.2 Method

### 4.2.1 Selection of Experimental points

Considering the above objectives, in Experiment 2 we assigned poor detection sensitivity ($d'_H = 1$) to all the participants and good detection sensitivity ($d'_A = 2.3$) to both classification systems. All participants examined 2 different classification systems that differed only in their preprogrammed response criterion. One system used a response criterion of $\beta_A = 1$, which matches a participants' optimal response criterion ("matching $\beta$"), while the other used $\beta_A = 0.03$, reflecting considerably different incentives than those of the participants ("Different $\beta$"). For the above settings, the ResQu model [Douer and Meyer 2020] predicts a theoretical responsibility value of 12% for the matching response criteria and 73% for the different response criteria.

The above two experimental points have the advantage of enabling meaningful comparison to the results of Experiment 1. First, the condition with matching $\beta$ is identical to an experimental point in Experiment 1, in which the measured responsibility of "less-accurate" participants deviated substantially from the optimal theoretical value when they used an "accurate" system. Hence, including this experimental point in Experiment 2 allowed us to examine whether this empirical deviation was replicated in another experiment. Secondly, the experimental point with different $\beta$ has a predicted theoretical responsibility of 73%, which is close to a theoretical responsibility of 69% that was predicted in Experiment 1 for identical human parameters ("less-accurate" human) but a different type of classification system ("less-accurate" with "matching $\beta$"). Hence, this experimental point enabled us to compare the empirical responsibility values in both cases, to examine if they are also similar (i.e. if similar theoretical responsibility values, due to different system parameters, lead to similar empirical responsibility values).

### 4.2.2 Participants

Participants were 30 students from Tel Aviv University (ages 20-34, median 25, 53% females), of which 28 were undergraduate students and 25 belonged to the Faculty of Engineering. The method of recruitment and the monetary reward for participation were the same as in Experiment 1.

### 4.2.3. Apparatus, Procedure and Design

The apparatus was identical to that of Experiment 1, and so were the prior information given to the participants, the participants' payoff scheme, the rate of defective items, the number of trials with each of the two classification systems, and the questionnaires.

In correspondence with the selected experimental points, all participants had a similar level of overlap between the two distributions of long and short rectangle population, which was set to represent the selected level of human detection sensitivity of $d'_H = 1$ ("less-accurate" human). All participants saw classification results from two "accurate" classification system, which had high detection sensitivity ($d'_A = 2.3$) but had different response criteria. One system used a response criterion of $\beta_A = 1$, which matches a participants' optimal response criterion, while the other used $\beta_A = 0.03$, reflecting considerably different incentives. We counterbalanced the order of the systems, with 15 participants first using the "matching $\beta$" system and then the "different $\beta$" system, while the other 15 used the systems in the reversed order. Table 8 summarizes the outcome probabilities of the two classification systems in the experiment and the systems' positive and negative predictive values.

TABLE 8. Outcome probabilities for the different classification systems in Experiment 2

| Type of System | Parameters | Defective (Signal) | | Intact (Noise) | | PPV | NPV |
|---|---|---|---|---|---|---|---|
| | | Red True Positive | Green False Negative | Red False Positive | Green True Negative | | |
| **Matching $\beta$** | $d'_A$=2.3, $\beta_A$=1 | 87% | 13% | 13% | 87% | 82% | 91% |
| **Different $\beta$** | $d'_A$=2.3, $\beta_A$ =0.03 | 99.6% | 0.4% | 65% | 35% | 51% | 99% |

The classification system with a matching $\beta$ had the same parameters and outcome probabilities as the accurate classification system in Experiment 1. The classification system with a different $\beta$ reflected a much higher incentive to reduce the acceptance of defective units (False Negatives). This came at the price of an increase in the rate of False Positives (false alarms). Due to the rate of defective items, whenever this classification system indicated an intact item, there was a 99% chance that it was indeed intact. On the other hand, this system's PPV was only 51%, so only about half of the items the system indicated as defective were indeed defective.

## 4.3. Results and Discussion

### 4.3.1. Measured Responsibility

The ResQu model predicts that when human detection sensitivity is poor and inferior to the high detection sensitivity of the classification system, the human should rely much less on a system that uses a considerably different response criterion (i.e. assume much higher responsibility) than on a system that uses a similar response criterion.

We analyzed the measured human responsibility with a two-way mixed analysis of variance (ANOVA), with the order in which the classification systems were examined as a between-subjects variable and the type of classification system as a within-subjects variable. As in Experiment 1, there were no significant effects of the order of experiencing the two classification systems. As predicted by the ResQu model, the changes in the response criterion between the two systems had a significant large effect $F(1,28)$= 110.43, $MSE$=2.7, $Par.\ \eta^2$= .80, $\eta^2_G$ = .67, $p$ < .0001, $Obs.\ Power$ =1.0. Participants assumed significantly higher measured responsibility with the classification system that used a different response criterion ($mean$= .85, $Sd$= .02), compared to the classification system that used a matching response criterion ($mean$= .43, $Sd$= .04).

When the system used a different response criterion, the measured responsibility was 85%, close to the theoretical prediction of 73%. This value is close to the theoretical responsibility of 69% the model predicted in Experiment 1 for identical human parameters but a different type of classification system (a "less-accurate" system with "matching $\beta$"). In Experiment 1 the corresponding measured responsibility was 77%. Hence, the empirical results, across the two experiments, are coherent. When humans with the same capabilities worked with two different classification systems (in terms of their detection sensitivity and response criterion), for which the theoretical model predicted similar responsibility values, the empirical measured responsibility was also similar.

When the system used a matching response criterion, the measured responsibility (43%) was very close to the one observed in the corresponding identical experimental point in Experiment 1 (46%), much higher than the theoretical prediction of 12%. Therefore, the empirical results, across the two experiments, are coherent as the empirical deviation from the optimum in Experiment 1 was replicated in Experiment 2.

We investigated the sources for deviations of the measured responsibility from the optimal theoretical value by analyzing the SDT measures $d'_{eff}$ and cutoff difference (see Table 9 for the optimal and mean empirical values). In both classification systems, the empirical $d'_{eff}$ was lower than the corresponding maximal value, implying that participants did not optimally utilize the information from the system. With both classification systems, the cutoff difference was lower than the corresponding optimal theoretical value, implying that participants tended to under-trust the two systems.

TABLE 9. Theoretical predictions Vs. empirical results

| System Response criterion $\beta$ | ResQu - Measured Responsibility | | | SDT - $d'_{eff}$ | | | SDT - Cutoffs Difference | | |
|---|---|---|---|---|---|---|---|---|---|
| | Theoretical Optimum | Empirical Mean | Diff. | Theoretical Optimum | Empirical Mean | Diff. | Theoretical Optimum | Empirical Mean | Diff. |
| Different $\beta$ ($\beta_A = 0.03$) | 73% | 85% | 12% | 1.4 | 1.0 | -0.4 | 5.0 | 0.7 | -4.3 |
| Matching $\beta$ ($\beta_A = 1$) | 12% | 43% | 31% | 2.3 | 2.0 | -0.3 | 3.9 | 2.0 | -1.9 |

Detection Sensitivities: Human $d'_H$=1; Sytem $d'_A$=2.3

At first, there seems to be an inconsistency between the deviations of the measured responsibility and the cutoff difference from their optimal values. The measured responsibility deviated more for the classification system with the matching response criterion, but the cutoff difference deviated more for the system with a different response criterion. The inconsistency may be explained by refining the analysis of the participants' behavior, by analyzing their levels of reliance and compliance. Table 10 summarizes the analysis.

TABLE 10. Theoretical SDT levels of reliance and compliance Vs. empirical result

| System | Different Response Criterion ($\beta$ =0.03) | | | Matching Response Criterion ($\beta$ =1) | | |
|---|---|---|---|---|---|---|
| | Reliance | Compliance | Trust | Reliance | Compliance | Trust |
| | Cutoff point for a Green indicator | Cutoff point for a Red indicator | Cutoff difference | Cutoff point for a Green indicator | Cutoff point for a Red indicator | Cutoff difference |
| Probability for indicator color | 20% | 80% | -- | 57% | 43% | -- |
| Theoretical cutoff | 4.6 | -0.4 | 5 | 1.95 | -1.95 | 3.9 |
| Empirical cutoff | 0.6 | -0.1 | 0.7 | 1 | -1 | 2 |
| Difference (Theory Vs Empiric) | 4 | -0.3 | 4.3 | 0.95 | -0.95 | 1.9 |

For the classification system with a matching response criterion, there were similar levels of under-reliance and under-compliance, so with both types of system indications, participants deviated quite substantially from the optimal theoretical value (by .95), leading to a cutoff difference of 1.9.

For the classification system with a different response criterion, the situation is quite different, as the optimal theoretical cutoff setting is asymmetrical (4.6 and -.4 respectively), due to a large difference between its NPV and PPV values (see Table 8). With this system, participants only deviated considerably from the optimal value when the system indicator was green (Difference = 4) but used an almost optimal value when it was red (Difference = -.3). Thus, the difference between the theoretical predictions and the empirical behavior is mainly due to participants' under-reliance. It is important to note that with this system, in 80% of the trials participants saw a red indication and therefore behaved close to the theoretical prediction. In this respect, the standard SDT measure of cutoff difference is misleading with this system, as it does not take into account the relative probabilities for the occurrence of reliance and compliance.

From the above we can see that with the classification system that used a different response criterion, in 80% of the trials the participants deviated only *slightly* from the optimal value (by .3), while they deviated quite considerably (by .95) in all trials with the system that used a matching response criterion. Therefore, in 80% of the trials, the participants deviated much more with the classification system that used a matching response criterion.

To conclude, after considering the relative probabilities for the occurrence of reliance and compliance, we see that there is consistency between the deviations of measured responsibility and the cutoff difference from the optimal values. The participants assumed much higher-than-optimal responsibility with the system that used a matching response criterion. The analysis, based on SDT measures of trust, showed that these participants overestimated their own capabilities, which led them to select of non-optimal cutoffs, due to both under-reliance and under-compliance.

The above analysis also demonstrates that the ResQu model's measure of responsibility has some advantages over SDT measures that are used to analyze human trust. The ResQu model's measure of responsibility, which is based on entropy, considers the different base probabilities for signals, system classification results and human responses, and reflects the share of unique human contribution to the outcomes in a single easily interpretable value. Conversely, the SDT measure of absolute cutoff difference might be misleading, as it does not consider the relative probabilities for the occurrence of reliance and compliance behaviors.

### 4.3.2. Subjective self-responsibility

Questions *Q3-Q5* referred to the participants' subjective assessments of their own responsibility. We reverse-scored questions *Q3* and *Q4* for the score to reflect responsibility. A reliability analysis, using Cronbach's $\alpha$ showed $\alpha = .82$. Hence we used the mean of the three questions as a measure for the subjective responsibility.

We analyzed the different types of subjective assessments with a two-way mixed analysis of variance (ANOVA), with the type of classification system as a within-subject variable and the order in which the systems were examined as a between-subject variable. There was no significant effect of the order of experiencing the two systems, so we focus on reporting the result regarding the effects of the system's response criteria.

Question *Q1* referred to the subjective assessment of the system's classification ability. The change of response criterion between the two classification systems had a significant large effect on the perceived capability of the systems $F_{(1,28)} = 57.29$, *MSE* $=77.07$, *Par.* $\eta^2 = .67$, $\eta^2_G = .55$, *Obs. Power* $=1$, $p < .0001$. The ability of the system with a matching response criterion was rated significantly higher (*mean* $= 5.9$, *Sd* $= .13$) than the system with a different response criterion (*mean* $= 3.6$, *Sd* $= .24$).

Question *Q2* referred to the subjective assessment of the participant's own classification ability. Here, the participants rightly perceived their own ability as being the same with both classification systems. The change of the response criterion between the two systems had no significant effect on the participants' perceived self-capabilities $F_{(1,28)} = .14$, *MSE* $= .15$, *Par.* $\eta^2 = .01$, $\eta^2_G = .00$, *Obs. Power* $= .07$, $p = .71$. The mean scores were 4.2 (*Sd* $= .21$) for the classification system with the different response criterion and 4.1 (*Sd* $= .23$) for the system with a matching response criterion.

The mean score for (reverse coded) questions *Q3, Q4* and question *Q5* reflected the subjective assessment of human responsibility. As predicted by the ResQu model, the change of response criterion between the two classification systems had a significant large effect on the perceived responsibility $F_{(1,28)} = 53.4$, *MSE* $= 54.78$, *Par.* $\eta^2 = .66$, $\eta^2_G = .47$, *Obs. Power* $= 1.00$, $p < .0001$. The participants assessed that they had much less responsibility with the system with the matching response criterion (*mean* $= 2.3$, *Sd* $= .16$) and much more responsibility with the system with the different response criterion (*mean* $= 4.2$, *Sd* $= .22$). This result is coherent with the subjective assessments expressed in *Q1* and *Q2* regarding the system and human performance.

The normalized average values of subjective responsibility were identical to those of the measured responsibility which implies that, as in Experiment 1, participants judged their true level of marginal contribution with each classification system quite accurately.

### 4.3.3. Measured responsibility and subjective assessments as mediators

In Experiment 2, we used manipulated proxy for theoretical responsibility, in the form of two different system response criteria, of which one reflected similar incentives to those of the human, and the other reflected a considerable difference.

As in the analysis of Experiment 1, we examined a hypothesis that the actual "hands on" experience with each classification system (i.e. the measured responsibility), served as a mediator to forming the subsequent subjective responsibility assessments. We first analyzed a simple mediation model to examine whether the direct effect of the manipulated proxy for theoretical responsibility on the subjective perception of responsibility (expressed in questions *Q3-Q5*) was partially or fully mediated by the measured responsibility. Fig. 10 summarizes the model results. All coefficients in the figure are fully standardized, and the confidence intervals are bias corrected bootstrapped CI based on 5000 samples.

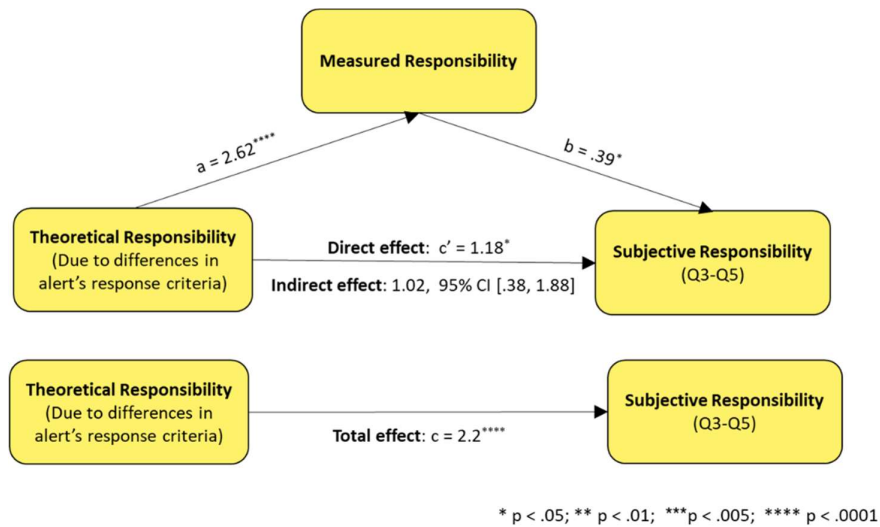

* p < .05; ** p < .01; ***p < .005; **** p < .0001

*Fig.* 10. Model of theoretical responsibility as a predictor of subjective responsibility, mediated by measured responsibility.

Fig. 10 reveals a positive and very significant effect of theoretical responsibility on both the measured and the subjective responsibilities. More importantly, the results show a significant indirect effect on subjective responsibility through the mediation of measured responsibility, *b*=1.02 (SE=.38), 95% BCa CI [.38,1.88]. The total effect of theoretical responsibility on subjective responsibility was divided almost equally between a direct and an indirect effect (with the measured responsibility as a mediator).

As in Experiment 1, the formation of subjective responsibility was not based mainly on the actual level of human contribution, but on other subjective factors. Such possible factors could be the separate preceding subjective assessments of system capabilities (*Q1*) and human capabilities (*Q2*). To examine this hypothesis, we expanded the mediation model to include parallel multiple mediation of these additional factors. Fig. 11 summarizes the results. All coefficients in the figure are fully standardized, and the confidence intervals are bias corrected bootstrapped CI based on 5000 samples.

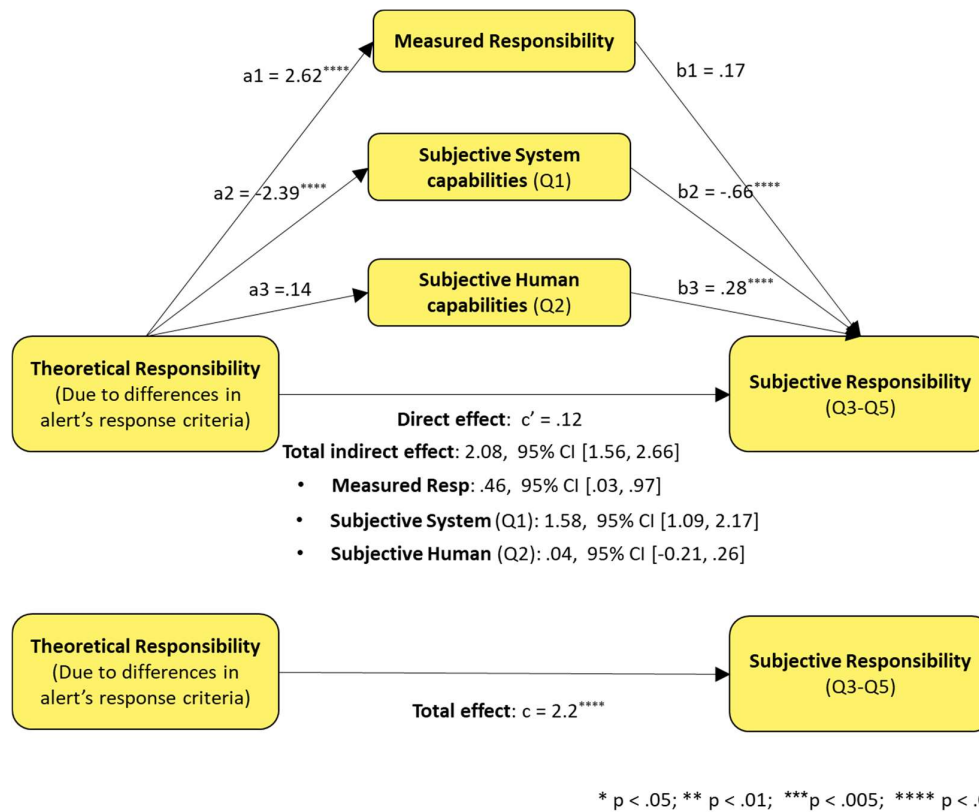

*Fig.* 11 Model of theoretical responsibility as a predictor of subjective responsibility, mediated by measured responsibility and subjective assessments of human and system capabilities.

As in Experiment 1, the results expose a large significant total indirect effect of theoretical responsibility on subjective responsibility through the mediation by both measured responsibility and the subjective assessments of system and human capabilities, $b$=2.08 (SE=.27), 95% BCa CI [1.56,2.66]. After inclusion of the additional mediators, the total indirect effect constituted most of the total effect (94%), while the direct effect was reduced to a small portion (6%). Most of the total indirect effect (75%) is due to large significant mediation via the subjective assessment regarding the system capabilities $b$=1.58 (SE=.27), 95% BCa CI [1.09,2.17]. The rest of the total indirect effect was due to significant mediation via measured responsibility $b$=.46 (SE=.24), 95% BCa CI [.03,.97]. The mediation effect via the subjective assessment of human capabilities was not significant $b$=.04 (SE=.12), 95% BCa CI [-.21,.26]

To conclude, as in Experiment 1, the above results expose a significant large indirect effect of theoretical responsibility on subjective responsibility via mediation. Again, the most influencing mediator was the subjective assessment of system capabilities. The measured responsibility (i.e. the actual level of human contribution), despite being also a significant mediator had less influence.

### 4.3.3. Subjective own responsibility vs. another person's responsibility

Question *Q6* referred to the subjective assessment of another person's responsibility. We analyzed the different subjective assessments with a two-way mixed analyses of variance (ANOVA), with the order of the systems as between-subjects variable and the type of the system as a within-subjects variable. There was no significant main effect of the order, nor any significant interaction that involved the order. The differences between the two types of systems were significant with a medium-large effect size $F(1,28)$= 9.75, *MSE*=19.27, *Par.* $\eta^2$= .26, $\eta^2_G$ = .15, *Obs. Power* =.85, *p* =.004. The accurate participants rated another person's responsibility with the accurate system significantly lower (*mean*= 4.0, *Sd*= .27) than with the less accurate system (*mean*= 5.2, *Sd*= .28).

The normalized average values of self-subjective responsibility were identical to those of another person's responsibility which implies that, like the results of Experiment 1, participants rated the responsibility of another person, in the same situation, similar to their own responsibility.

## 5.3. Conclusions for Experiment 2

Experiment 2 examined the effect of threshold differences when human detection sensitivity is poor and inferior to the high detection sensitivity of the system:

(1) As predicted by the ResQu model, the threshold difference between the human and the system had a significant effect on the measured and subjective responsibility. The human relied much less on the classification system that used a considerably different response criterion, taking on more responsibility with that system, despite its far better detection abilities

(2) When participants used the classification system with a different response criterion, they assumed higher measured responsibility, which was close to the predicted theoretical value. However, when participants used the classification system with the same response criterion as their own, the measured responsibility deviated from the optimal theoretical value, to a similar degree as in Experiment 1. An analysis, based on SDT measures of trust, showed that these participants overestimated their own capabilities, which led them to select non-optimal cutoffs, due to both under-reliance and under-compliance.

(3) A comparison between measured and subjective responsibilities showed that participants subjectively judged their actual contribution with each classification system quite accurately. There was a significant effect of theoretical responsibility (manipulated through different system response criteria) on both measured and subjective responsibilities. However, the effect on subjective responsibility was largely indirect, via mediation. The most influencing mediator was the subjective assessment of system capabilities. The measured responsibility (i.e. the actual level of human contribution), despite being also a significant mediator, had less influence.

(4) Participants rated the responsibility of another person, in the same situation, similar to their own responsibility.

## 5 DISCUSSION AND FUTURE WORK

### 5.1 Main Results

In two laboratory experiments, participants performed an aided classification task, in which we controlled the classification abilities and incentives of both humans and systems. We compared the theoretical predictions of a newly developed responsibility quantification model to the actual measured responsibility participants took on and to their subjective rankings of responsibility.

As predicted by the ResQu model, when the human and system use the same threshold, their relative classification abilities (i.e. their relative detection sensitivities) had a significant effect on both the measured and subjective responsibility. It decreased as the system's classification ability improved and increased as the human classification ability improved. When the incentives differed, participants in the poor classification ability condition relied much less on a system with superior classification ability but a considerably different threshold. When using it, they assumed much higher responsibility, despite the system's far better classification abilities, confirming a non-intuitive prediction of the theoretical model.

*Measured Responsibility*: The ResQu model generally provided quite accurate predictions of the actual values of measured responsibility. In most cases, the mean value of the measured responsibility was close to the optimal theoretical value predicted by the ResQu model. Thus, there was a strong correlation between the mean values of measured responsibility, across the six experimental groups, and their corresponding theoretical prediction (Pearson correlation coefficient of $r=0.98$, $p<.0001$, a nonparametric Spearman correlation coefficient $r_s=0.97$, $p=.001$). The theoretical and the mean measured responsibility only differed to a large extent when participants with poor capabilities used a system with much superior capabilities. In line with previous results from behavioral research, these participants overestimated their own capabilities, relied less-than-optimally on the system and, thus, assumed greater-than-optimal responsibility.

*Subjective Responsibility*: Participants judged their marginal contribution with each system quite accurately. Consequently, there was a strong correlation between the mean values of measured responsibility, in the six experimental groups, and their corresponding mean values of subjective responsibility scores (Pearson correlation coefficient $r=0.943$, $p=.005$; a nonparametric Spearman correlation coefficient $r_s=0.900$, $p=.01$). As predicted, there was a significant effect of theoretical responsibility on subjective responsibility. However, this effect was largely indirect, via mediation by other variables. The most influencing mediator was the subjective assessment of system capabilities. Measured responsibility was also a significant mediator, but with a weaker effect. This implies that the subjective assessment of self-contribution was mainly determined by perceptions of the quality of the system and less by observing the actual contribution when using the system. Lastly, participants subjectively rated the responsibility of another person, in the same situation, similar to their own responsibility.

The above result demonstrate that the ResQu model is not merely a theoretical model. It is also a descriptive model that allows us to predict human's measured and subjective responsibility in interactions with intelligent systems and advanced automation. One can consider the characteristics of the human, the system and the environment, as well as systematic behavioral deviations. The ResQu model value can be used to assess the actual human responsibility or how humans will perceive their own responsibility. In addition, the ResQu measure of responsibility reflects the share of unique human contribution in a single easily interpretable value, and thus has some advantages over other SDT measures that are used to analyze human compliance and reliance in aided decision making. Hence, the ResQu model is well suited to serve as a new measure for quantifying user trust and involvement in intelligent systems.

## 5.2 Discussion

System designers are often required to keep humans in the loop to supervise systems and to handle unexpected events, even when the human may have limited abilities to do so. Our results demonstrate that, with advanced intelligent systems, simply putting a human into the loop does not assure that the human will have a meaningful role and unique contribution to the outcomes.

A prominent example for that are provisions regarding automated decision making in the EU General Data Protection Regulation (GDPR), which says in Article 22(1): "The data subject shall have the right not to be subject to a decision based solely on automated processing, including profiling, which produces legal effects concerning him or her or similarly significantly affects him or her." The interpretation of this article states that it constitutes a general prohibition for decision making, based solely on automated processing. This means that to conform to the GDPR requirements, institutions will have to involve humans in automated decision processes. Realistically speaking, these humans will, of course, base their decisions largely on information from decision support systems. Thus, despite the seeming adherence to the requirement not to fully automate the decision process, the actual comparative human responsibility, according to the ResQu model, will likely be minimal. Here, again, organization should consider the true added value of the human to system processes, beyond the simple role of approving system decisions.

Another key example concerns the developments of autonomous weapon systems (AWSs), which in the event of a failure, can cause catastrophic consequences, such as mass fratricide or civilian casualties. The rapid technological developments in AWSs have raised concerns that humans will have limited ability to supervise or be involved in their use, to prevent occurrences of such adverse consequences (Scharre 2016). As with other intelligent systems, this leads to a responsibility gap in the ability to divide causal responsibility between the human operator and the AWS (Gerdes 2018).

The concerns prompted requests to restrict or even ban the development of advanced AWSs (Asaro 2012, Goose 2015, Guersenzvaig 2018, Noorman 2014, Noorman and Johnson 2014, Walsh 2015). Governments respond to these concerns with the assurance that a human will always be kept in the loop, whenever an AWS exerts lethal force. For example, the explicit policy of the U.S. Department of Defense is that these systems "shall be designed to allow commanders and operators to exercise appropriate levels of human judgment over the use of force" (USDD 2012)]. The UK policy is that the operation of weapon systems will always be under human control, and that no offensive systems should be able to prosecute targets without involving a human (ICRC, 2014). However, near future AWS technologies will almost certainly outperform humans in many critical tasks, such as ability to assess the likelihood of harming civilians while engaging a selected target, distinguishing between combatants and non-combatants, and making decisions and acting in uncertain environments which require very short reaction times [Douer & Meyer 2020]. Hence, again, the simplistic regulatory demands to keep a human in the loop in advanced AWSs can be misleading and even futile, as the comparative human casual responsibility will likely be minimal. A well-known example [Horowitz & Scharre 2015] clarifies this point by describing an AWS operator who sits in an isolated room, pressing a button every time a light goes on, which authorizes the firing of a weapon. Here the human is kept in the loop, actively authorizing the AWS's action selection, so there is seemingly compliance with the regulatory demands. But with no additional self-information, the human control is far from being meaningful and the comparative human responsibility is insignificant. Moreover, as we have seen in our analysis of empirical responsibility, operators of advanced AWSs may feel (correctly) that they have no significant impact on the system and outcomes, and may attempt to be more involved by interfering more than necessary or conversely feel less motivated to take necessary actions [Hassenzahl & Klapperich 2014, Rangarajan, et al. 2005, Smith et al. 1999]. In general, both responses will be obstacles to exploiting the full performance potential of the system, but, more severely, in AWSs this can lead to fatal consequences that could have otherwise been prevented.

One needs to be aware of the demands to keep humans in the loop and to facilitate meaningful human control. One must be prepared to deal with their implications on the overall functioning of the system and on the humans' attitudes towards the system and their role in it. The ResQu model enables system designers to identify such cases in advance and to take them into consideration when planning the human role in the system (e.g. by assigning the human additional duties which are more meaningful) or when negotiating system requirements with regulators.

Our results also showed that when human operators interact with advanced intelligent systems with capabilities that greatly exceed those of the humans, they will almost necessarily have only limited comparative causal responsibility for the outcomes. This may create discrepancies between their *role* responsibility, which describes the duties of the human operators they are held accountable for, and their *causal* responsibility, which describes their actual unique contribution to the outcomes. There are two possible sources for this inconsistency. First, operators may lack the authority to take the necessary actions to fulfill their role, and thus may have limited ability to influence the system outcomes. This is the known responsibility-authority double bind [Woods 1985, Woods 2004, Pritchett, Kim, & Feigh 2014].

The ResQu model exposes an additional source for discrepancy. Since causal responsibility is measured in respect to the outcomes, it is influenced by uncertainties and probabilistic aspects that are not part of authority (which is defined explicitly and granted beforehand). Thus, the human operator may be granted sufficient authority, but due to system design and probabilistic factors related to the environment and the system, the operator's decision and actions have only minor marginal influence on the outcomes.

One example are autonomous cars. They may have the "final word" in steering and braking in order to avoid an accident, preceding or even overriding the human steering or braking actions when there is a need to act in much shorter time than common human reaction times. In addition, in certain extreme environmental conditions, such as dark nights, fog, or sandstorms, the human driver may depend almost exclusively on information from the car's advanced multiple sensors. In such cases, the human driver's ability to supervise the car's systems will be reduced. The driver contributes little to the outcomes, despite having both the role and authority to do so. The ResQu model enables system designers and regulators to identify such discrepancies between causal and role responsibilities in advance and to take them into consideration.

Our results revealed that subjective assessments of responsibility were mainly mediated via the subjective assessment of system capabilities and to lesser extent via the actual human contribution or the perception of self-abilities. Hence, one can provide the human with continuous feedback regarding the system and human performance (e.g. measures such as precision and recall values in a classification task) and the level of marginal unique contribution (e.g. the ResQu responsibility score). This can help the human operators and outside observers, such as managers, better calibrate subjective assessments of responsibility.

In the interaction with intelligent systems, the human may be considered fully legally responsible for adverse outcomes, even without sufficient control to prevent them or when contributing very little to create these outcomes. With the advent of advanced intelligent systems, with abilities that clearly exceed those of humans in many critical functions, a choice will have to be made. One can progress to autonomous systems with very little human involvement, or alternatively maintain a certain level of human involvement at the price of lowering system performance. The decision whether one wants to sacrifice the full potential of system performance to increase human involvement and responsibility will have to be made on a case-by-case basis by regulators and system designers. The insights gained from using the ResQu model can support this process. The intermediate option, where systems will be increasingly intelligent, while still keeping the human in the loop, can possibly lead to the inclusion of humans to simply fulfill regulatory demands without them having any real added value for system performance. Falsely claiming that the human is responsible for adverse outcomes may expose her or him to a psychological burden of self-blaming or legal liability, even when the human contributed very little to create those outcomes.

The three types of responsibility measures we introduced in the current study can potentially be used in legal procedures and the formation of regulation, by exposing anomalies and providing a new method for quantifying the actual human comparative casual responsibility for the outcomes. Each measure may be used in a different context. The theoretical responsibility measure is most appropriate when one wants to specify the optimal level of human involvement in a system. Nevertheless, it quantifies the responsibility of a perfectly rational human, and as such, it can be used as a theoretical benchmark for optimal behavior. However, it does not necessarily provide adequate descriptions of what humans will actually do. The measured responsibility may be the best predictor of how the 'average person' behaves when interacting with a system. Finally, the subjective responsibility reflects the impressions people involved in using the systems have regarding their marginal level of contribution. Thus, the combined analysis of these three responsibility measures can perhaps help in preparing suitable regulations and legal treatment regarding human roles and responsibility in the interaction with intelligent systems. This can be done by tying different design options to their predicted effects on the users' behavior and perceptions of responsibility.

## 5.3 Limitations and Future Work

The study was conducted in a controlled lab environment, using a simple, abstract experimental setting, in which participants performed a simple task and received immediate feedback on their performance. This experimental setting allowed us to control the relevant independent variables and to compare the results to the theoretical responsibility predictions of the ResQu model, while still capturing central properties of human interactions with decision support systems. Nevertheless, the lab experiment may not fully represent complex human interactions with intelligent systems in real-world environments. Thus, future work should expand the research by applying the ResQu model to real-world settings.

Also, the empirical study was limited to Israeli students, of which most were undergraduate students from the faculty of engineering. This sample population may not fully reflect cultural or educational effects on the degree to which people take and judge responsibility. Future work should examine whether, and to what degree, cultural and educational differences affect measured and subjective responsibility.

Based on measures of information theory, the ResQu model assumes that the combined human-machine system is stationary and ergodic. Hence it is not directly applicable when human characteristics are not stationary (e.g. when there are learning or fatigue effects that lead to a change in the level of human involvement over time). In this case one can calculate the responsibility measure repeatedly, at different points in time, and look if convergence exists.

It is important to note that the three responsibility measures focus on estimating the mean share of unique human contribution to the outcomes, averaged over the distributions of possible states in the environment, system performance and human responses. As such, they do not deal with the retrospective evaluation of human responsibility in a specific single past event. Such an analysis deals with *retrospective* responsibility (in contrast to the *prospective* analysis we present in our paper). Retrospective responsibility is very important in the contexts of identifying causal sequences leading to outcomes, finding fault and perusing legal justice, and it is an important topic for future work.

Future work should also address temporal effects, such as the time required to make a decision and its implications on the human's tendency to rely on the system and on the corresponding measures of theoretical, measured and subjective responsibility. We plan to address this issue, too, in future work.

# 6 CONCLUSIONS

Intelligent systems have become ubiquitous and are major parts of our life. As they become more advanced, the human comparative causal responsibility for outcomes becomes equivocal.

In this paper, we analyzed the descriptive abilities of a newly developed theoretical model of human responsibility to predict actual human behavior and perceptions of responsibility in laboratory interactions with an automated decision aid. In two lab experiments we compared the theoretical responsibility values to the actual measured responsibility a person took on and to the subjectively perceived responsibility, for various combinations of human and system characteristics.

Our results showed that the ResQu model is not only a theoretical model, but also a descriptive model that allows us to predict human's measured and subjective responsibility. One can consider the characteristics of the human, the system and the environment, as well as systematic behavioral deviations, and can compute a ResQu value that can serve to assess the actual human responsibility or how humans will perceive their own responsibility.

Our study has far-reaching implication. When humans interact with advanced intelligent systems with capabilities that greatly exceed their own, it is almost inevitable that the comparative unique human contribution (i.e. ResQu responsibility) will be small, even if formally the human is included in the process. Simply putting a human into the loop does not assure that the human will play a meaningful role. This may create discrepancies between humans' *role* responsibility and their *causal* responsibility, and may also expose them to unjustified legal measures and psychological burdens. In addition, humans that interact with such advanced intelligent systems may feel (correctly) that they have no significant impact on the system and outcomes and attempt to be more involved by interfering more than necessary. Conversely, they may feel less motivated to take necessary actions. Both responses will probably limit the ability to exploit the full potential of the system.

We do not prescribe, recommend or criticize different ways to involve humans in a process. Rather, our model describes human involvement in processes and quantifies the comparative causal responsibility of the human for outcomes, given the properties of the situation. This analysis should be part of the evaluation of different system design alternatives and the relevant regulations. A specific design may give humans a more central role in a process, but this may come at the prize of limiting the use of the capabilities the intelligent systems may have to offer. Similarly, introducing advanced AI capabilities into an existing system will almost necessarily lower human casual responsibility and involvement. One needs to be aware of these changes and be prepared to deal with the implications they may have on the functioning of the system and on the humans' attitudes towards the system and their role in it.